# Cogwheel phase cycling in population-detected optical coherent multidimensional spectroscopy

Ajay Jayachandran[†,1], Stefan Mueller[†,1], and Tobias Brixner[1,2,3]*

[1]*Institut für Physikalische und Theoretische Chemie, Universität Würzburg, Am Hubland, 97074 Würzburg, Germany*
[2]*Center for Nanosystems Chemistry (CNC), Universität Würzburg, Theodor-Boveri-Weg, 97074 Würzburg, Germany*
[3]*Institute for Sustainable Chemistry & Catalysis with Boron (ICB), Universität Würzburg, Am Hubland, 97074 Würzburg, Germany*

*†These authors contributed equally*
*\*Electronic mail: tobias.brixner@uni-wuerzburg.de*

## Abstract

An integral procedure in every coherent multidimensional spectroscopy experiment is to suppress undesired background signals. For that purpose, one can employ a particular phase-matching geometry or phase cycling, a procedure that was adapted from nuclear magnetic resonance (NMR) spectroscopy. In optical multidimensional spectroscopy, phase cycling has been usually carried out in a "nested" fashion, where pulse phases are incremented sequentially with linearly spaced increments. Another phase-cycling approach which was developed for NMR spectroscopy is "cogwheel phase cycling," where all pulse phases are varied simultaneously in increments defined by so-called "winding numbers". Here we explore the concept of cogwheel phase cycling in the context of population-based coherent multidimensional spectroscopy. We derive selection rules for resolving and extracting fourth-order and higher-order nonlinear signals by cogwheel phase cycling and describe how to perform a numerical search for the winding numbers for various population-detected 2D spectroscopy experiments. We also provide an expression for a numerical search for nested phase-cycling schemes and predict the most economical schemes of both approaches for a wide range of nonlinear signals. The signal selectivity of the technique is demonstrated experimentally by acquiring rephasing and nonrephasing fourth-order signals of a laser dye by both phase-cycling approaches. We find that individual nonlinear signal contributions are, in most cases, captured with fewer steps by cogwheel phase cycling compared to nested phase cycling.

# I. Introduction

Coherent multidimensional optical spectroscopy has served as an excellent tool for studying a wide range of chemical systems including light-harvesting systems,[1–4] molecular dye aggregates,[5–9] perovskite semiconductors,[10,11] quantum dots,[12–17] and others. The additional excitation frequency axis compared to traditional one-dimensional optical spectroscopy techniques such as transient absorption spectroscopy,[18,19] impulsive stimulated Raman spectroscopy,[20] or fluorescence-detected pump–probe spectroscopy,[21,22] helps in disentangling congested spectral features by separating overlapping spectral features and resolving inhomogeneous broadening of spectral line shapes.[23,24] Coherent two-dimensional (2D) spectroscopy is among the most widely explored multidimensional techniques utilizing different wavelength regions of the optical spectrum.[25–30] Since the inception of 2D spectroscopy in the 1990s,[31,32] numerous detection methods and geometries have been developed to receive information on static and dynamical properties of a system under study. The nonlinear response of the system can be obtained either through a coherently detected scheme or a population-based detection scheme ("action detection"). The associated experiments can be performed in a noncollinear, partially collinear, or fully collinear geometry, utilizing either phase matching, phase cycling, or phase modulation, which can be viewed as "dynamic phase cycling," for capturing the desired nonlinear signals.[27,33–46]

Using phase matching, various nonlinear signals are spatially separated along different phase-matching directions given by the linear combination of different wavevectors $\mathbf{k}_r$ ($r$ = 1, 2, 3, 4) corresponding to each pulse $r$ of a four-pulse sequence. In pump–probe geometry, higher-order multiquantum signals with the same phase-matching signature can also be separated by resolving these signals spectrally along the excitation axis. Phase modulation is another approach to separate nonlinear signals. This technique uses nested interferometers which split a single pulse into a pulse train with controllable interpulse delays. In every interferometer arm, acousto-optic modulators (AOM) modulate each pulse of a four-pulse sequence with a unique radio frequency $\Omega_r$ ($r$ = 1, 2, 3, 4). Similar to phase matching, a linear combination of these frequency components helps to extract different nonlinear signals. In phase modulation, the carrier-envelope phase is frequency-shifted at an individual $\Omega_r$ without affecting the laser repetition rate. The nonlinear signals with their unique phase signature oscillate at a frequency equivalent to the linear combinations of $\Omega_{12}$ and $\Omega_{34}$ due to this dynamic phase cycling and the resulting signal contributions are demodulated on-the-fly using conventional or digital lock-in.[35,47–50] While phase modulation continuously modulates the phase difference between pulses, phase cycling is based on systematic incrementation of interpulse phases, which can be carried out in a shot-to-shot basis using a pulse shaper.[42,43]

Phase cycling traces its roots to nuclear magnetic resonance (NMR) spectroscopy and has been adopted into 2D optical spectroscopy.[42,51–55] The close correlation between 2D NMR and its optical analogs and the significance of adopting phase cycling has been identified previously in works by researchers like David Jonas.[32,56,51,57–59,40] The main purpose of phase cycling is to suppress unwanted signals through destructive interference and, at the same time, to generate constructive interference for desired signals.[60]

The traditional formulation of phase cycling, referred to as nested phase cycling,[52,61] has so far been most commonly employed in both coherently detected or action-detected multidimensional spectroscopies.[40,62,63] In nested phase cycling, only one absolute pulse phase is varied at a time while the phases of the other pulses are kept constant before incrementing the absolute phase of another pulse. Another variant of phase cycling introduced in NMR spectroscopy is so-called "cogwheel phase cycling" which was proposed and demonstrated by Levitt and coworkers.[64] In that approach, all pulse phases are incremented simultaneously. In the work by Levitt et al., cogwheel phase cycling was found to lead to fewer phase-cycling steps in 2D NMR spectroscopy experiments for separating nonlinear signals without aliasing and it was concluded that cogwheel phase cycling is more economical than nested phase cycling.[64,65] Recently, we have experimentally verified the applicability of this phase-cycling technique in fluorescence-detected 2D spectroscopy by exemplarily measuring the two-quantum photon-echo signal in a pulse-shaper-based three-pulse experiment.[53] Generally, it is desirable to keep the number of scanning steps as low as possible, especially to enable one to perform highly dimensional nonlinear spectroscopies, i.e., with more than three frequency dimensions, in a reasonable amount of time. Such highly dimensional methods fully resolve a high-order nonlinear response over more multiple frequency dimensions with maximum spectral resolution.[66] Moreover, a shorter total measurement time is required if the sample of interest shows a strong photobleaching behavior or exhibits an otherwise limited chemical stability. If targeted nonlinear signals could be acquired in less time, the time saved could be used to further average data to increase data quality. It is therefore worth determining whether and to what extent measurement time reductions are possible with cogwheel phase cycling in optical spectroscopy methods.

In this work, we describe the concept of cogwheel phase cycling in optical population-based multidimensional spectroscopy. We establish cogwheel selection rules for action-detected three-pulse and four-pulse experiments in a collinear beam geometry. We provide the most economic cogwheel phase-cycling schemes for extracting nonlinear signal contributions of different orders of nonlinearity. Experimentally, we demonstrate the signal selectivity of cogwheel phase cycling through four-pulse population-detected multidimensional experiments using fluorescence as the detected observable.

## II. Cogwheel and nested phase cycling

### A. Definition of quantities

Before we explain the general differences between cogwheel and nested phase cycling, we introduce our nomenclature of the population-based nonlinear signals. For that purpose, we classify the various nonlinear signals similarly as in NMR spectroscopy, that is, by the coherence orders induced by the phase-specific excitation sequence. In NMR spectroscopy, the coherence order $p_{ji} \in \mathbb{Z}$ describes the type of coherence that evolves between two radio-frequency pulses $i$ and $j$, where $j > i$. In particular, the coherence order is a quantum number attributed to a superposition between spin eigenstates that differ in their total spin magnetic quantum number by $p$.[60,67] In optical electronic spectroscopy, the concept of coherence orders can be directly transferred to electronic excitations. Hence, we label each nonlinear signal with the respective absolute order of a coherence between electronic states that evolves during a certain time interval between the optical excitation pulses, followed by a "Q" for "quantum".[68] For example, a coherence that oscillates with a frequency within the range of frequencies given by the laser spectrum is referred to as a one-quantum (1Q) coherence. Coherences that oscillate with a frequency within the second and third harmonic of the laser spectrum are called two-quantum (2Q) and three-quantum (3Q) coherences, respectively. Oscillatory dynamics that evolve with frequencies below those contained in the fundamental laser spectrum are categorized as zero-quantum (0Q) coherences. Note that we also classify non-oscillatory population dynamics as "0Q". We further distinguish between rephasing (R) and nonrephasing (NR) contributions for a signal with given coherence types.[69] In a rephasing signal, the coherences evolving over two different time delays are phase conjugate whereas they are not in a nonrephasing signal.[69] In a three-pulse experiment with time delays $\tau$ and $t$, [Fig. 1(a), left], there are only two coherence types $XQ$ and $ZQ$ with $X, Z \geq 1$ leading to the label "$XQ$–$ZQ$".[55] In a four-pulse experiment [Fig. 1(a), right], the coherence time delays remain the same; however, there is a third delay $T$ in between, leading to the label "$XQ$–$YQ$–$ZQ$" with $Y \geq 0$. Note that, in general, $X$, $Y$, and $Z$ can all be $\geq 0$, however, we will only focus on those signals that can be extracted by phase cycling without aliasing (assuming that nonlinear orders higher than the desired one do not contribute). As an example of aliasing, consider the phase signature $-\varphi_1 + \varphi_1 + \varphi_2 + \varphi_3$ of a desired fourth-order 0Q–1Q signal acquired by three pulses. This phase signature coincides with the second-order 0Q–1Q signal as the first pulse effectively acts with zero phase. Thus, the fourth-order 0Q–1Q signal is aliased with a lower-order signal and cannot be unambiguously extracted from a three-pulse 2D experiment unless one employs an additional "intensity cycling" procedure.[70]

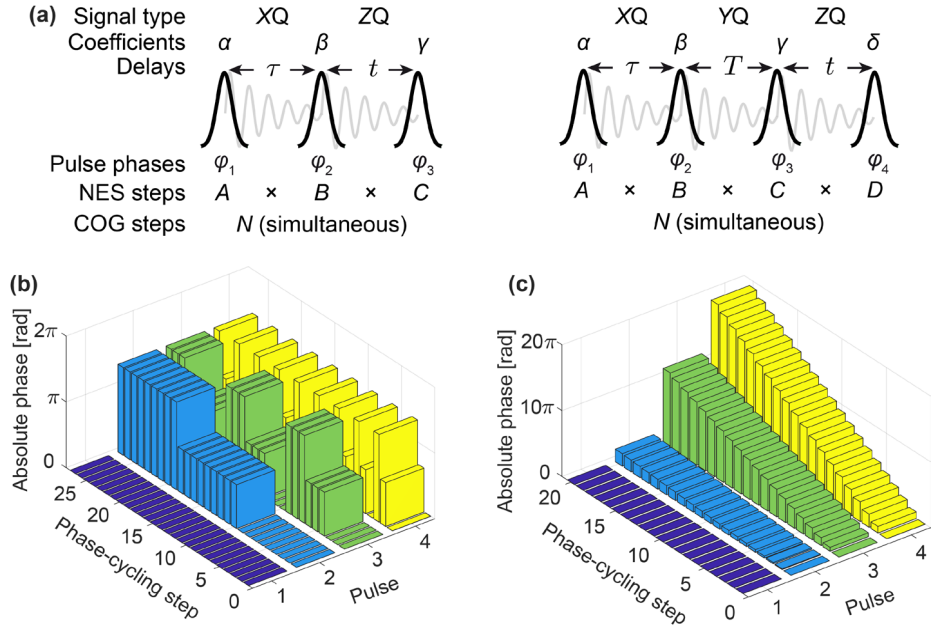

**Fig. 1.** Definition of population-based nonlinear signals and quantities in nested and cogwheel phase cycling. (a) Schemes of a three-pulse sequence with time delays $\tau$ and $t$ (left) and a four-pulse sequence with delays $\tau$, $T$, and $t$ (right). Nonlinear signals are described by the absolute coherence orders $X$, $Y$, and $Z$ (where "Q" denotes "quantum") and a set of phase coefficients $\alpha$, $\beta$, $\gamma$, and $\delta$ of the individual pulse phases $\varphi_1$, $\varphi_2$, $\varphi_3$, and $\varphi_4$, respectively. In nested (NES) phase cycling, pulse phases are incremented sequentially according to a *A×B×C×D* scheme whereas in cogwheel (COG) phase cycling, all pulse phases are varied simultaneously, but with pulse-specific increments. (b) Visualization of 27-fold (1×3×3×3) nested phase cycling and (c) 20-fold cogwheel phase-cycling "COG20(0, 1, 6, 10)".

Each nonlinear signal is further associated with a unique total phase given by a linear combination of the absolute phases $\varphi_i$ ($i$ = 1, 2, 3, …) of the individual pulses which is

$$\varphi_{XQ\text{-}ZQ} = \alpha\varphi_1 + \beta\varphi_2 + \gamma\varphi_3 \tag{1}$$

in case of a three-pulse sequence and

$$\varphi_{XQ\text{-}YQ\text{-}ZQ} = \alpha\varphi_1 + \beta\varphi_2 + \gamma\varphi_3 + \delta\varphi_4 \tag{2}$$

when a four-pulse sequence is considered. The coefficients $\alpha$, $\beta$, $\gamma$, and $\delta$ ($\alpha, \beta, \gamma, \delta \in \mathbb{Z}$) describe the multiples and the signs of the interactions imprinted by the pulse phases $\varphi_1$, $\varphi_2$, $\varphi_3$, and $\varphi_4$ on the system density matrix.[52]

In the conventional "nested" phase-cycling approach, the pulse phases are varied independently and sequentially in discrete steps. In other words, only the phase of one pulse is varied at a time while the phases of the other pulses are kept constant. This leads to the pulse phases

$$\begin{aligned}\varphi_1 &= \frac{2\pi}{A}m_1, \\ \varphi_2 &= \frac{2\pi}{B}m_2, \\ \varphi_3 &= \frac{2\pi}{C}m_3, \\ \varphi_4 &= \frac{2\pi}{D}m_4,\end{aligned} \tag{3}$$

where *A*, *B*, *C*, and *D* are the numbers of the phase-cycling steps of the individual pulses ($A \times B \times C \times D = N$) and $m_i$ is the phase-cycle counter of pulse $i$ ($m_1$ = 0, 1, 2, …, $A$ – 1; $m_2$ = 0, 1, 2, … $B$ – 1; $m_3$ = 0, 1, 2, …, $C$ – 1; $m_4$ = 0, 1, 2, …, $D$ – 1). Thus, the $m_i$ value of a particular pulse is varied once at a time while those $m_i$ of the other pulses are kept constant such that all combinations of the possible values of {$m_1$, $m_2$, $m_3$, $m_4$} are sampled in a nested phase-cycling scheme.

The situation is different in cogwheel phase cycling. Here, the pulse phases are varied simultaneously in discrete steps, but with pulse-specific increment sizes that are specified by so-called winding numbers. Levitt and coworkers describe this scheme metaphorically as a "set of meshing cogwheels" with "a defined ratio of angular velocities".[64] In that approach, the pulse phases are defined as

$$\varphi_1 = \frac{2\pi w_1}{N} m,$$

$$\varphi_2 = \frac{2\pi w_2}{N} m,$$

$$\varphi_3 = \frac{2\pi w_3}{N} m,$$

$$\varphi_4 = \frac{2\pi w_4}{N} m,$$

(4)

with the pulse-specific winding numbers $w_i \in \{0, 1, 2, 3, \ldots\}$, the total number of phase-cycling steps $N$, and the mutual phase-cycle counter $m$ ($m = 0, 1, 2, \ldots, N-1$).[65]

To further illustrate the differences between the two phase-cycling approaches, we exemplarily show plots of a 27-fold phase-cycling scheme [Fig. 1(b)] and an equivalent 20-fold cogwheel phase-cycling scheme [Fig. 1(c)]. Both approaches facilitate the acquisition of the rephasing 1Q–0Q–1Q signal contribution which will be further discussed in Section III.

Analogous to Ref. [52], one can also consider inter-pulse phases, i.e., the phase of the first pulse is considered as constant, e.g., zero, and the phases of the other pulses are referenced to the first pulse ($\varphi_{i1} = \varphi_i - \varphi_1$ with $i = 1, 2, 3, 4$). If the phases are referenced in this way, the winding number differences $w_{i1} = w_i - w_1$ are also referenced to the first pulse ($w_{i1} \in \mathbb{Z}$).

### B. Cogwheel phase cycling for three-pulse experiments

In NMR spectroscopy, the concept of coherence orders is used to classify the various signals and to define selection rules for phase-cycling approaches. For three-pulse NMR experiments, there are two coherence orders $p_{21}$ and $p_{32}$. The selection rule for finding a suitable cogwheel phase-cycling scheme for a desired signal with coherence orders $p_{21}^0$ and $p_{32}^0$ derived by Levitt et al. is

$$[w_{21}(p_{21} - p_{21}^0) + w_{32}(p_{32} - p_{32}^0)] \bmod N \begin{cases} = 0 \text{ for the desired signal} \\ \neq 0 \text{ for undesired signals,} \end{cases}$$

(5)

with $w_{21} = w_2 - w_1$ and $w_{32} = w_3 - w_2$ as the winding number differences between the first two and the last two pulses, respectively.[64] One can show that the coherence orders can be directly related to the pulse-specific phase coefficients $\alpha$, $\beta$, and $\gamma$ used in population-based multidimensional spectroscopy.[53] In particular, the coherence order $p_{21}$ is equivalent to the coefficient $\alpha$ since $p_{21}$ is solely dependent on the phase imprinted by the first pulse. Likewise, $p_{32}$ is a result of the interaction pattern of the first two pulses, thus $p_{32} = \alpha + \beta$. Using the general rule for population-based signals,[52]

$$\alpha + \beta + \gamma = 0,$$

(6)

which must be obeyed to reach a population state after the interaction with the excitation sequence, the coherence orders can thus be expressed as

$$p_{21} = -\beta - \gamma,$$
$$p_{32} = -\gamma. \tag{7}$$

The selection rule for population-based signals then follows from Eqs. (5)–(7),

$$[w_1(\alpha_0 - \alpha) + w_2(\beta_0 - \beta) + w_3(\gamma_0 - \gamma)] \bmod N \begin{cases} = 0 \text{ for the desired signal} \\ \neq 0 \text{ for undesired signals,} \end{cases} \tag{8}$$

with $\alpha_0$, $\beta_0$, and $\gamma_0$ as the coefficients of the desired signal. If the phases of the second and third pulse are referenced to that of the first pulse, Eq. (8) can also be expressed as[53]

$$[w_{21}(\beta_0 - \beta) + w_{31}(\gamma_0 - \gamma)] \bmod N \begin{cases} = 0 \text{ for the desired signal} \\ \neq 0 \text{ for undesired signals.} \end{cases} \tag{9}$$

Note that because of the restriction of Eq. (6), not every set of $\alpha$, $\beta$, and $\gamma$ corresponds to a measurable signal which somewhat narrows down the search for suitable winding numbers.

A suitable cogwheel phase-cycling scheme exhibits winding numbers which ensure that only the desired signal is acquired while all the other signals up to the order of nonlinearity $n$ of the desired signal are fully suppressed. Note that the selection rules of Eqs. (8) and (9) only hold under the assumption that nonlinear signals which have a higher order than that of the desired signal do not contribute,[52] i.e.,

$$|\alpha| + |\beta| + |\gamma| \leq n. \tag{10}$$

If higher orders than $n$ are a significant part of the response, the lower-order signals become contaminated by them and one must seek additional, more elaborate procedures for the suppression of these contaminations such as so-called "intensity cycling".[68,70] The solutions of Eq. (8) represent the cogwheel phase-cycling schemes which can be written, in the original notation, "COG$N(w_1, w_2, w_3)$," or "COG$N(0, w_{21}, w_{31})$," whereby in the latter, the pulse phases are referenced to the first pulse.[64] In the case of NMR spectroscopy, it has been shown that adequate solutions can be found efficiently using a computer algorithm.[65]

Similar selection rules can also be formulated for nested phase cycling. To this end, by using Eq. (3) and Eq. (4), the winding numbers are simply expressed as $w_1 = N/A$, $w_2 = N/B$, and $w_3 = N/C$, whereas $N = A \times B \times C$. Since all pulse phases are varied independently and not simultaneously, it follows from Eq. (8) that a set of selection rules must be fulfilled simultaneously for finding a suitable nested phase-cycling scheme:

$$\begin{aligned} & [BC(\alpha_0 - \alpha)] \bmod N \\ \wedge & [AC(\beta_0 - \beta)] \bmod N \\ \wedge & [AB(\gamma_0 - \gamma)] \bmod N \end{aligned} \begin{cases} = 0 \text{ for the desired signal} \\ \neq 0 \text{ for undesired signals.} \end{cases} \tag{11}$$

For the remainder of the paper, we thus use the notation "NES$N(A, B, C)$" for all nested three-pulse phase-cycling schemes (which are also commonly denoted as "$A \times B \times C$").

Using the selection rules of Eq. (8) and Eq. (11), we can determine the most economical cogwheel phase-cycling schemes and nested phase-cycling schemes, respectively, needed to acquire various population-based nonlinear signal contributions. For that purpose, we implemented a respective numerical search algorithm in Matlab.[71] To define appropriate termination conditions, we searched for winding numbers in a range of $-15 \leq w_{i1} \leq 15$ and a maximum number of phase-cycling steps of $N_{max} = 30$. Note that, for the numerical search for the winding numbers, search time can be saved using Eq. (9), since only two winding number differences $w_{i1}$ [instead of three winding numbers as in Eq. (8)] have to be determined. For each signal, we list an exemplary minimum-step cogwheel phase-cycling scheme in Table 1. Note that there are various other cogwheel phase-cycling schemes possible;[64] however, as an arbitrary selection criterion, we list in Table 1 only those schemes for which the sum of the absolute values of the winding numbers is the smallest. Given by the chosen termination conditions, phase-cycling schemes with much larger phase increment size, i.e., the magnitude of the winding number relative to the total number of phase cycling steps, are generally possible solutions to Eq. (9) as well.

**Table 1.** Most economical cogwheel (COG) and nested (NES) phase-cycling schemes for population-based three-pulse experiments. Note that we display for each signal only one out of many possible cogwheel phase-cycling schemes. The label "R" denotes rephasing signals and "NR" denotes nonrephasing signals.

| Order | Signal type | $\tau$ | $t$ | Minimal phase-cycling schemes | |
|---|---|---|---|---|---|
| | | | | Cogwheel | Nested |
| 4 | R | 1Q | 1Q | COG10(0, 1, 5) | NES10(1, 5, 2) |
| | NR | 2Q | 1Q | COG10(0, −4, 1) | NES10(5, 1, 2) |
| | NR | 1Q | 2Q | COG10(0, 5, 1) | NES10(1, 2, 5) |
| 6 | R | 2Q | 1Q | COG18(0, 3, 1) | NES21(1, 7, 3) |
| | R | 1Q | 2Q | COG18(0, −2, 1) | NES21(1, 7, 3) |
| | NR | 3Q | 1Q | COG18(0, 3, 2) | NES21(7, 1, 3) |
| | NR | 3Q | 2Q | COG18(0, 2, 3) | NES21(7, 1, 3) |
| | NR | 2Q | 3Q | COG18(0, 1, 3) | NES21(1, 3, 7) |
| | NR | 1Q | 3Q | COG18(0, −1, 2) | NES21(1, 3, 7) |
| 8 | R | 2Q | 2Q | COG27(0, 1, 9) | NES27(1, 9, 3) |
| | NR | 4Q | 1Q | COG24(0, 3, 2) | NES36(9, 1, 4) |
| | NR | 4Q | 2Q | COG27(0, −8, 1) | NES27(9, 1, 3) |
| | NR | 4Q | 3Q | COG24(0, 2, 3) | NES36(9, 1, 4) |
| | NR | 1Q | 4Q | COG24(0, 1, 3) | NES36(1, 9, 4) |
| | NR | 2Q | 4Q | COG27(0, 9, 1) | NES27(1, 3, 9) |
| | NR | 3Q | 4Q | COG24(0, −1, 2) | NES36(1, 4, 9) |

From Table 1, it is evident that cogwheel phase-cycling requires fewer steps in many cases, especially for signals higher than fourth order, which in turn leads to time savings in experiments. However, no time savings can be achieved for fourth-order signals where both nested and cogwheel phase cycling require at least 10 steps. While the number of required phase-cycling steps can be reduced by 14% for all sixth-order signals, reductions up to 33% can be achieved for eighth-order signals. On the other hand, there are eighth-order signals where both nested and cogwheel phase cycling require the same number of steps, like for the rephasing 2Q–2Q signal,[53] for example.

To extract the desired complex-valued nonlinear signal $\tilde{S}$ from the phase- and delay-dependent raw data $S$ of a population-based 2D experiment with nested phase cycling, a suitable linear combination of the raw data has to be constructed according to[52]

$$\tilde{S}(\tau, t, \beta_0, \gamma_0) = \frac{1}{BC} \sum_{b=0}^{B-1} \sum_{c=0}^{C-1} S(\tau, t, b\varphi_{21}, c\varphi_{31}) \exp(-ib\beta_0\varphi_{21})\exp(-ic\gamma_0\varphi_{31}). \quad (12)$$

If cogwheel phase cycling is carried out, then Eq. (12) changes to

$$\tilde{S}(\tau, t, \beta_0, \gamma_0) = \frac{1}{N} \sum_{m=0}^{N-1} S(\tau, t, m\varphi_{21}, m\varphi_{31}) \exp[-im(\beta_0\varphi_{21} + \gamma_0\varphi_{31})]. \quad (13)$$

Figure 2 illustrates the equivalence of information obtained from both phase-cycling approaches. As an example, we simulated the rephasing 2Q–1Q signal, in which a 2Q coherence is rephased by means of a 1Q coherence [Fig. 2(a)].[55,72] The linewidth along the antidiagonal thus represents the convolution of the homogeneous dephasing times of 1Q and 2Q coherences of the system. The simulation was carried out by using our open-source Matlab toolbox QDT,[73] which explicitly calculates the propagation of the system density matrix via the Lindblad master equation.[74] As a model system, we used the same parameters as in a previous publication;[73] however, we incorporated an additional doubly excited state with a population relaxation time of 30 fs [see Fig. 2(b) for an energy-level scheme]. The pure dephasing times associated with coherences between the doubly excited state |3⟩ and the other states were set to 60 fs. The laser excitation spectrum was modelled as Gaussian with an energy of 1.95 eV and thus covers the transitions that connect the ground state |0⟩ with the singly excited states |1⟩ and |2⟩ as well as the singly excited states with |3⟩ at the same intensity. All transitions have the same dipole moment $\mu_{fi}$ (with $i$ as the initial and $f$ as the final state). We performed a simulation with the minimal NES21(1, 7, 3) scheme and the more economical COG18(0, 3, 1) scheme, both by scanning $\tau$ and $t$ from 0 to 90 fs with 31 steps each in a fully rotating frame.

The real parts of the 2D spectra obtained by nested and cogwheel phase cycling are shown in Fig. 2(c) and 2(d), respectively. Both 2D spectra show two positively signed peaks corresponding to the correlation of the two singly excited states with the doubly excited state. It is immediately clear that the amplitude, sign, and line shape of the two 2D spectra are virtually identical, which proves that the two different phase-cycling approaches yield the same information content, with cogwheel phase cycling requiring 14% fewer steps. Upon subtraction of the 2D spectrum calculated with COG18 from that calculated with NES21, we obtain a difference spectrum in Fig. 2(e). Note that the maximum amplitude of this difference spectrum is about three orders of magnitude smaller than that of the 2D spectra calculated with COG18 and NES21.

We have also carried out additional simulations where we increased the electric field amplitude (see Sec. SI in the supplementary material). Concluding, we find indications that cogwheel phase cycling may be more susceptible to higher-order contaminations than nested phase cycling. However, we also find that the difference in the degree of contamination in the cases we studied has no significant relevance for experimental data, as these differences are at a signal level that is comparable to the experimental noise level (see also Sec. SII in the supplementary material).

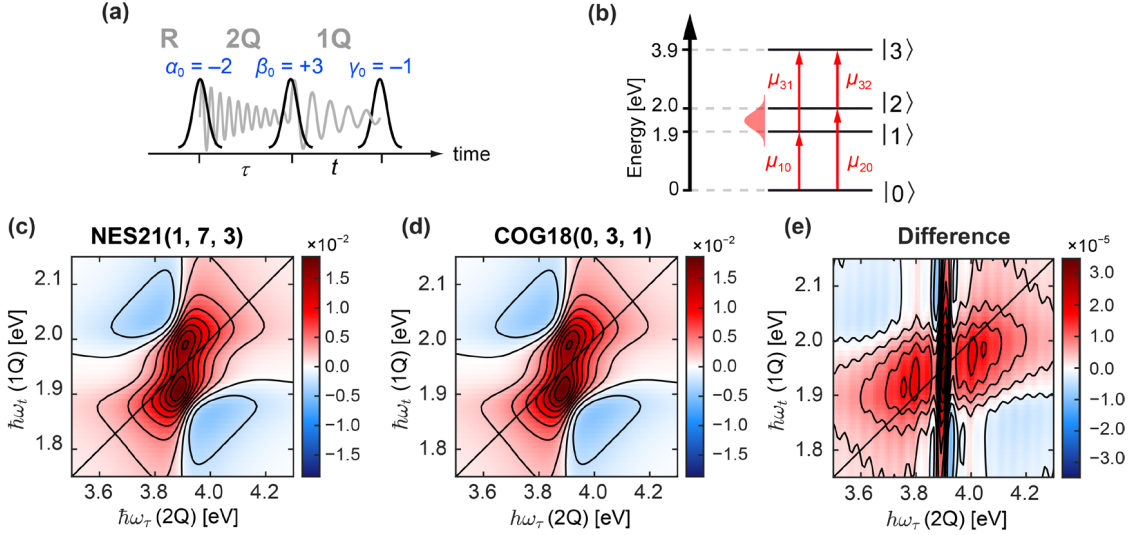

**Fig. 2.** Cogwheel phase cycling versus nested phase cycling in the simulation of an exemplary three-pulse experiment. (a) Pulse sequence (black) and signal phase coefficients (blue) of the rephasing 2Q–1Q signal with probed coherences depicted in gray. (b) Energy level scheme of the model system with ground (0), singly excited (1 and 2) and doubly excited (3) states which are connected by transition dipole moments $\mu_{fi}$, where $f$ denotes the final and $i$ the initial state, and which all have the same magnitude (red arrows). The laser spectrum (red Gaussian) covers the transition energies of the 0–1 and the 0–2 transitions equally. (c) Real part of the rephasing 2Q–1Q 2D spectrum in case of 21-fold nested phase cycling and (d) 18-fold cogwheel phase cycling. (e) Difference spectrum, obtained from subtracting the spectrum of (d) from that of (c). All 2D spectra are drawn with nine linearly spaced contour lines.

## C. Cogwheel phase cycling for four-pulse experiments

Having defined the selection rules for three-pulse experiments, we can proceed analogously for four-pulse experiments. In a four-pulse sequence, we consider the additional winding number difference $w_{41}$ and the pulse-phase specific coefficient $\delta$. The selection rule for suitable cogwheel phase-cycling schemes is then

$$[w_{21}(\beta_0 - \beta) + w_{31}(\gamma_0 - \gamma) + w_{41}(\delta_0 - \delta)] \bmod N \begin{cases} = 0 \text{ for the desired signal} \\ \neq 0 \text{ for undesired signals.} \end{cases} \quad (14)$$

In general, the selection rule for an experiment with $R$ pulses can be written as[60]

$$\left[\sum_{i=1}^{R}(\vartheta_{i,0} - \vartheta_i)w_{i1}\right] \bmod N \begin{cases} = 0 \text{ for the desired signal} \\ \neq 0 \text{ for undesired signals} \end{cases} \quad (15)$$

with $\vartheta_i$ and $\vartheta_{i,0}$ as the pulse-specific phase coefficients where the subscript zero denotes the coefficients of the desired signal.

For nested phase cycling, Eq. (11) is extended to

$$\left.\begin{array}{l}[BCD(\alpha_0 - \alpha)] \bmod N \\ \wedge\ [ACD(\beta_0 - \beta)] \bmod N \\ \wedge\ [ABD(\gamma_0 - \gamma)] \bmod N \\ \wedge\ [ABC(\delta_0 - \delta)] \bmod N\end{array}\right\} \begin{array}{l}= 0 \text{ for the desired signal} \\ \neq 0 \text{ for undesired signals.}\end{array} \quad (16)$$

The most economical cogwheel and nested phase-cycling schemes for four-pulse experiments are listed in Table 2. Note that also here, multiple solutions for suitable winding numbers are possible. For example, the fourth-order rephasing 1Q–0Q–1Q signal can be acquired not only by COG20(0, −9, 1, −5) but also by COG20(0, −14, −9, −10), COG20(0, 1, 6, 10), or COG20(0, 3, −7, 15). As for the three-pulse cogwheel phase-cycling schemes, in Table 2 we also only list the scheme with both the smallest number of steps and the smallest sum of the absolute values of the winding numbers. As termination conditions, we chose $-15 \leq w_{i1} \leq 15$, $N_{max} = 30$ for fourth-order signals, $N_{max} = 50$ for sixth-order signals, and $N_{max} = 100$ for eighth-order signals.

**Table 2.** Most economical cogwheel (COG) and nested (NES) phase-cycling schemes for action-detected four-pulse experiments. Note that for each signal we display only one out of many possible cogwheel phase-cycling schemes. Signals that probe 0Q coherences over $T$ can be classified as either rephasing ("R") or nonrephasing ("NR").

| Order | Signal type | $\tau$ | $T$ | $t$ | Minimal phase-cycling schemes Cogwheel | Nested |
|---|---|---|---|---|---|---|
| 4 | R | 1Q | 0Q | 1Q | COG20(0, −9, 1, −5) | NES27(1, 3, 3, 3) |
|  | NR | 1Q | 0Q | 1Q | COG20(0, −9, −5, 1) | NES27(1, 3, 3, 3) |
|  |  | 1Q | 2Q | 1Q | COG20(0, −10, −4, 1) | NES27(1, 3, 3, 3) |
| 6 | R | 2Q | 0Q | 1Q | COG45(0, 8, 5, 1) | NES60(5, 4, 3, 1) |
|  | NR | 2Q | 0Q | 1Q | COG45(0, 8, 1, 5) | NES60(4, 5, 3, 1) |
|  | R | 1Q | 0Q | 2Q | COG45(0, −2, −6, 3) | NES60(1, 3, 4, 5) |
|  | NR | 1Q | 0Q | 2Q | COG45(0, −2, 3, −6) | NES60(1, 3, 5, 4) |
|  |  | 1Q | 1Q | 1Q | COG45(0, −6, 3, 2) | NES60(1, 4, 5, 3) |
|  |  | 2Q | 1Q | 1Q | COG45(0, 5, 8, 1) | NES60(5, 3, 4, 1) |
|  |  | 1Q | 1Q | 2Q | COG45(0, −6, −2, 3) | NES60(1, 4, 3, 5) |
|  |  | 2Q | 1Q | 2Q | COG45(0, 5, 1, 8) | NES60(5, 3, 1, 4) |
|  |  | 1Q | 3Q | 1Q | COG45(0, −4, 5, 2) | NES60(1, 5, 4, 3) |
|  |  | 2Q | 3Q | 1Q | COG45(0, 1, 8, 5) | NES60(5, 1, 4, 3) |
|  |  | 1Q | 3Q | 2Q | COG45(0, −4, 2, 5) | NES60(1, 5, 3, 4) |
|  |  | 2Q | 3Q | 2Q | COG45(0, 1, 5, 8) | NES60(5, 1, 3, 4) |
| 8 | R | 2Q | 0Q | 2Q | COG81(0, −14, 13, −9) | NES125(1, 5, 5, 5) |
|  | NR | 2Q | 0Q | 2Q | COG81(0, −14, −9, 13) | NES125(1, 5, 5, 5) |

We have previously shown that all fourth- and sixth-order signals of Table 2 can be simultaneously acquired by a NES125(1, 5, 5, 5) phase-cycling scheme.[75] We identified that NES125(1, 5, 5, 5) is also the lowest-step scheme to resolve eighth-order signals such as the rephasing and nonrephasing 2Q–0Q–2Q signal contributions. These two contributions can be used to construct purely absorptive 2Q–0Q–2Q signals, which represent the direct 2Q analogue of the widely known absorptive 1Q–0Q–1Q signal.[35,50,54] If one is only interested in a specific signal contribution, however, selecting the appropriate cogwheel phase-cycling scheme requires fewer steps. While in three-pulse experiments, there are cases in which cogwheel phase cycling does not offer a reduction of the required phase-cycling steps compared to nested phase cycling (see Table 1), reductions can be consistently achieved in all kinds of four-pulse experiments. While measurement time can be reduced by 26% for all four-pulse fourth-order signals, all sixth-order signals can be acquired 25% faster as well. For eighth-order signals such as the rephasing and nonrephasing 2Q–0Q–2Q signals, an acquisition time reduction of 35% can be achieved. The eighth-order 2Q–0Q–2Q signal is expected to be helpful for measuring biexciton binding energies, tracking spectral diffusion of biexcitons, and the identification of coupling between different biexciton states within the manifold of excitonic states. In turn, this will help us understand intricate details of the electronic structure of molecular aggregates and semiconductor materials, for example.[76–79] We experimentally demonstrate the applicability of cogwheel phase cycling in four-pulse experiments on the example of two fourth-order signals in Section III. Similar to Eq. (13), the extraction of a desired signal acquired by cogwheel phase cycling is achieved by creating suitable linear combinations according to

$$\tilde{S}(\tau, T, t, \beta_0, \gamma_0, \delta_0) = \frac{1}{N} \sum_{m=0}^{N-1} S(\tau, T, t, m\varphi_{21}, m\varphi_{31}, m\varphi_{41}) \qquad (17)$$
$$\times \exp[-im(\beta_0 \varphi_{21} + \gamma_0 \varphi_{31} + \delta_0 \varphi_{41})].$$

On the basis of the data from Tables 1 and 2, a general trend can be derived: the ratio between the most economical cogwheel and nested phase-cycling steps decreases both with increasing number of required pulses and with increasing nonlinear order of a desired signal. Thus, it can be expected that even larger measurement time reductions may result for highly nonlinear signals with even higher dimensionality, i.e., signals which are acquired by more than four pulses and which are at least of sixth or higher order of nonlinearity. Ivchenko et al. have achieved a massive acquisition time reduction in NMR experiments involving five π pulses from 243 nested phase-cycling steps down to just eleven cogwheel phase-cycling steps.[80] To test whether analogous action-detected experiments may benefit from such acquisition time reductions, we determined a suitable cogwheel phase-cycling scheme for an exemplary action-detected signal that could be acquired using six optical pulses. By using a randomized search for the winding numbers in an interval $-50 \leq w_{i1} \leq 50$, we found that, e.g., COG132(0, 20, 30, 38, −36, 5) is sufficient to acquire a signal with phase signature $-\varphi_1 + \varphi_2 + \varphi_3 - \varphi_4 + \varphi_5 - \varphi_6$ which almost halves the acquisition time with regard to an otherwise required 243-step nested scheme.

# III. Experimental demonstration: collinear four-pulse experiment

## A. Setup description

Ultrafast laser pulses were generated by a Ti:sapphire regenerative amplifier (Spitfire Pro, Spectra-Physics, 800 nm, 35 fs pulse duration) with a repetition rate of 1 kHz. The output beam was attenuated by a beam splitter to ~1 W. This output beam was used to produce a white-light continuum by pumping a hollow-core fiber (Ultrafast Innovations GmbH) filled with argon (0.9 bar pressure).[81,82] The broadband output of the hollow-core fiber was compressed using a set of chirped mirrors (Ultrafast Innovations GmbH). The near-IR part along with the fundamental (800 nm) of the continuum was removed using a 750 nm short-pass filter (FESH750, Thorlabs). The beam size was adjusted using a telescope setup consisting of two focusing mirrors, and the beam was guided through an acousto-optic programmable dispersive filter (Dazzler, Fastlite). We used a dual grism compressor (Fastlite) to precompensate the chirp introduced by the pulse shaper which produced trains of four pulses with inter-pulse delays $\tau$, $T$ and $t$ in a shot-to-shot fashion in a fully rotating frame. We used the pulse shaper to cycle the pulse phases according to the framework of nested phase cycling or cogwheel phase cycling discussed in Section II. The output beam of the pulse shaper was then focused into a capillary flow cuvette (250 μm × 250 μm cross-section, Hellma) through which the sample was pumped by a microannular gear pump (mzr-2942-cy, HNP Mikrosysteme GmbH).

The sample consisted of the laser dye rhodamine 700 perchlorate (Radiant Dyes Laser Acc. GmbH) in ethanol (spectroscopy grade, Sigma-Aldrich) with 0.23 OD absorbance. The sample was excited using pulses with ~13 fs pulse duration (determined via pulse-shaper-assisted collinear frequency-resolved optical gating) and an energy of ~100 nJ (measured at maximum constructive interference of all four excitation pulses). The resulting fluorescence signal was detected using an avalanche photodiode (A-Cube S500-3, Laser Components) following its collection and recollimation using a pair of microscope objectives (04OAS010, CVI Melles Griot). The recollimated fluorescence signal was attenuated to a linear detector response level by neutral density filters (Newport) and then guided through an optical fiber (QP400-2-SR, Ocean Optics) which was connected to the photodiode.

The coherence delays $\tau$ and $t$ were scanned from 0 to 84 fs with a step size of 6 fs while the delay $T$ was scanned from 0 to 600 fs with a step size of 20 fs. To improve the signal-to-noise ratio, we averaged over 50 full data sets for all our measurements. After creating suitable linear combinations of the phase-dependent raw data to extract desired nonlinear signals, the time-domain data was apodized using a Hanning window and then five-fold zero-padded before 2D Fourier transformation.

## B. Experimental results

The chosen laser dye rhodamine 700 has its absorption maximum around 1.91 eV corresponding to the transition from the ground electronic state to the first excited electronic state. The absorption spectrum further features a minor vibronic shoulder at 2.08 eV in ethanol. We used a broadband laser spectrum centered around 1.90 eV to selectively excite the absorption maximum (Fig. 3) and collect the fluorescence centered around 1.85 eV.[83]

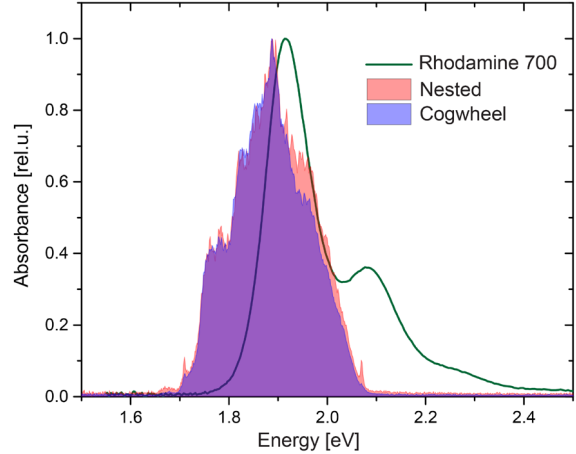

**Fig. 3.** The laser spectra for the NES27(1, 3, 3, 3) measurement (shaded orange area) as well as the COG20(0, 1, 6, 10) and COG20(0, 1, 10, 6) measurements (shaded purple area) are shown with the linear absorption spectrum of rhodamine 700 (green).

The most widely used signal in multidimensional optical spectroscopy is the fully absorptive third-order signal[57,84] which correlates 1Q coherences along both the excitation and detection axes. The respective analogue in an action-detected experiment is the fourth-order 1Q–0Q–1Q signal. Phase cycling enables us to extract both the rephasing and nonrephasing 1Q–0Q–1Q signals by weighting the raw data captured in the experiment with the signal-specific phase coefficients. For the rephasing and the nonrephasing signals, the coefficients are $\alpha_0 = -1, \beta_0 = 1, \gamma_0 = 1, \delta_0 = -1$ and $\alpha_0 = 1, \beta_0 = -1, \gamma_0 = 1, \delta_0 = -1$, respectively.[52]

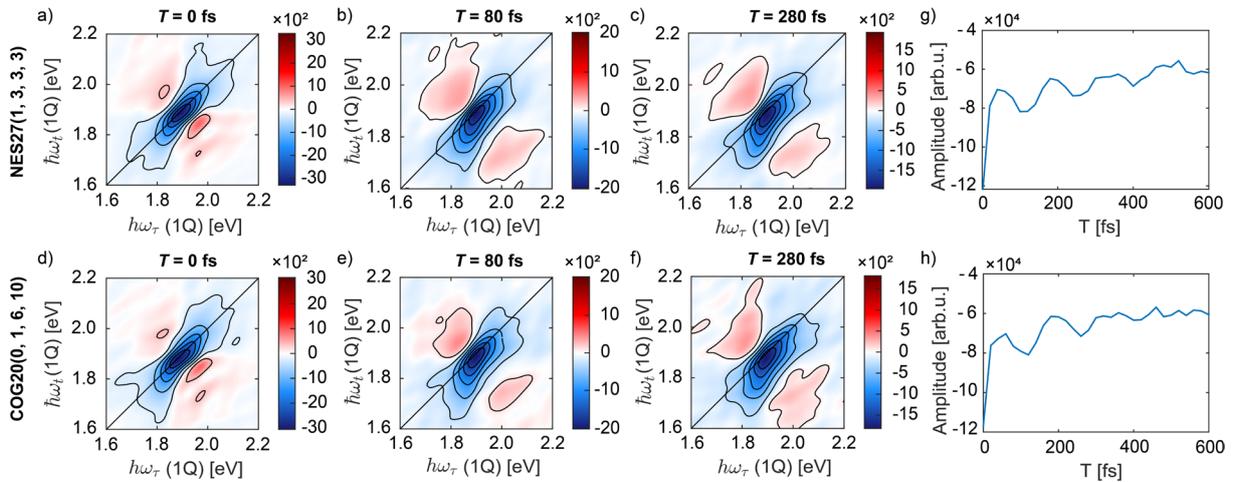

**Fig. 4.** Real parts of the rephasing 1Q–1Q 2D spectra of rhodamine 700 obtained for different $T$ delays. The upper row shows 2D spectra acquired by NES27(1, 3, 3, 3) at (a) $T = 0$ fs, (b) $T = 80$ fs, and (c) $T = 280$ fs. The lower row displays 2D spectra acquired by COG20(0, 1, 6, 10) at (d) $T = 0$ fs, (e) $T = 80$ fs, and (f) $T = 280$ fs. Integrating the signal amplitude within a region of interest (a square between 1.864 eV and 1.920 eV, not shown) for each $T$ yields oscillations with a period of 140 fs in both (g) NES27(1, 3, 3, 3) and (h) COG20(0, 1, 6, 10). All 2D spectra are drawn with six linearly spaced contour lines.

The signal selectivity of cogwheel phase cycling for the rephasing 1Q–0Q–1Q signal in a four-pulse experiment was verified using the COG20(0, 1, 6, 10) scheme while comparing it to the minimal nested NES27(1, 3, 3, 3) scheme (Fig. 4). Corresponding difference spectra, obtained by subtracting the spectra acquired by cogwheel phase-cycling from that acquired by nested phase cycling are shown in Fig. S2(a) in the supplementary material. As mentioned in Section II, COG20(0, 1, 6, 10) is one of many possible schemes involving the lowest number of cycling steps which avoids aliasing of the rephasing 1Q–0Q–1Q signal with other nonlinear signals.

In Fig. 4, it can be seen that we obtain similar features in terms of peak position and line shapes in the corresponding 1Q–1Q 2D spectra at various $T$ acquired by both nested and cogwheel phase cycling. The main diagonal peak is centered at 1.9 eV which closely matches the absorption maximum of the dye. The dispersive contribution of the nonlinear response is also present in form of phase twists, i.e., opposite-signed features on either side of the main peak with lower amplitude. With increasing $T$, there is a slight downward shift and broadening of the diagonal peak which can be associated with the Stokes shift known among the family of xanthene dyes to which rhodamine 700 belongs [Fig. 4(a)–(f)].[85]

Furthermore, by integrating the 2D spectra within the region of interest centered around the maximum signal amplitude (a square between 1.864 eV and 1.920 eV) for each $T$ delay and then plotting it against the corresponding time step, we see vibrational oscillations with a period of approximately 140 fs in both datasets recorded by nested and cogwheel phase cycling. These oscillations can be associated with the dominant vibrational modes of 231 cm$^{-1}$ and 252 cm$^{-1}$ of rhodamine 700 [Fig. 4(g) and 4(h)].[86]

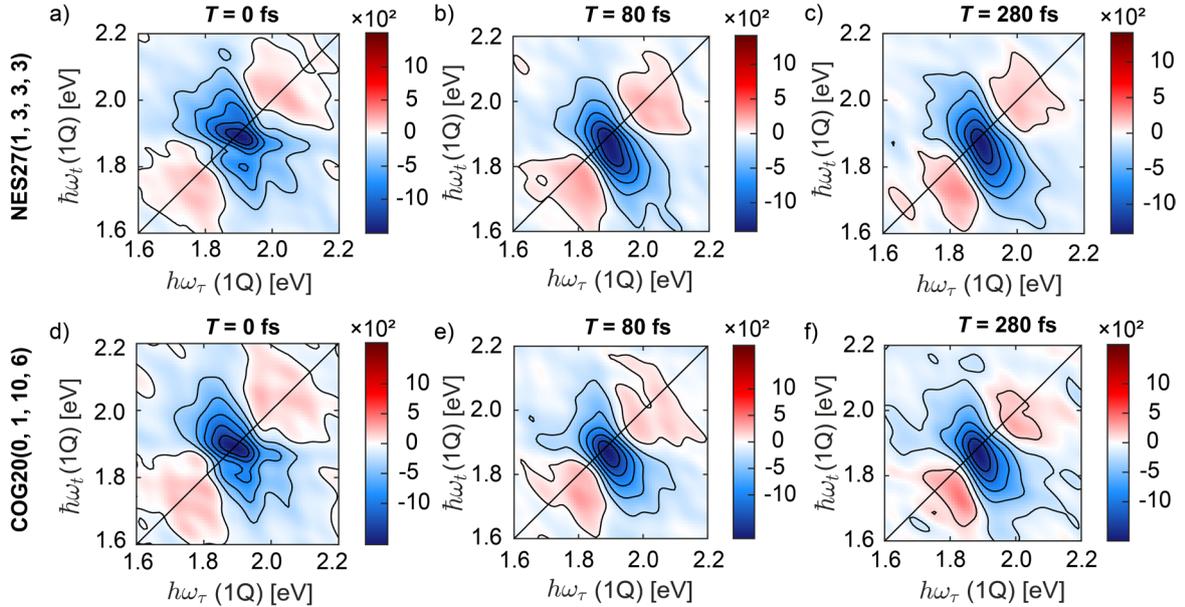

**Fig. 5.** Real parts of the nonrephasing 1Q–1Q 2D spectra of rhodamine 700 obtained for different $T$ delays. The upper row shows 2D spectra acquired by NES27(1, 3, 3, 3) at (a) $T = 0$ fs, (b) $T = 80$ fs, and (c) $T = 280$ fs. The lower row shows 2D spectra acquired by COG20(0, 1, 10, 6) at (d) $T = 0$ fs, (e) $T = 80$ fs, and (f) $T = 280$ fs. All 2D spectra are drawn with six linearly spaced contour lines.

In another experiment, we measured the nonrephasing component of the 1Q–0Q–1Q signal via cogwheel phase cycling. While the signal can be acquired simultaneously together with the rephasing 1Q–0Q–1Q signal with NES27(1, 3, 3, 3), we used COG20(0, 1, 10, 6) to measure it without aliasing. Difference spectra are reported in Fig. S2(b) in the supplementary material. Figure 5 shows that also for the case of the nonrephasing signal, we obtain the same information with both phase-cycling approaches. Again, fewer steps are required in cogwheel phase cycling. While the nonrephasing signal is smaller than the rephasing signal and hence has a somewhat lower signal-to-noise ratio because of the constructive interference via echo formation,[69] it is still the case that the 2D spectra measured by both approaches are consistent in terms of amplitude, line shape, and line-shape tilt for all measured $T$ delays. In both approaches, the main peak shifts to lower energies along the $\omega_t$ axis with increasing $T$. In addition, the line-shape tilt changes on the same time scale, indicating a relaxation process associated with the electronically excited state that occurs within 20–40 fs. Such line-shape changes can originate from spectral diffusion, vibrational relaxation, and the dynamic Stokes shift.[87–89]

## IV. Conclusion

We delineated the theory and selection rules for cogwheel phase cycling in three-pulse and four-pulse multidimensional optical spectroscopy with population-based detection. We provided the selection rules for determining the most economical cogwheel phase-cycling schemes and predicted the required schemes for several population-based signals up to eighth order of nonlinearity. By comparing the rephasing and nonrephasing 1Q–0Q–1Q signals acquired by nested and cogwheel phase cycling, we exemplarily demonstrated that cogwheel phase cycling provides the same information with fewer required measurement steps. While a 27-fold nested phase-cycling scheme is the minimum to acquire a rephasing or nonrephasing signal, each of these signals can be acquired by 20-fold cogwheel phase cycling, which results in a 26% reduction of acquisition time in that case. If only a specific nonlinear signal is desired, cogwheel phase cycling is, in most cases, more economical compared to nested phase cycling. Based on the derived selection rules, we showed that this is the case for several signals recorded by a three-pulse sequence. Even greater time savings can be achieved for signals that require a four-pulse excitation sequence. In general, we observed that the time savings achieved by cogwheel phase cycling increase with the number of pulses in the excitation sequence and with the order of nonlinearity of the desired signal. For eighth-order signals acquired by four pulses, for example, an acquisition time reduction of 35% can be achieved. The resulting reduction in acquisition time can be used to perform more averaging, which further improves the signal-to-noise ratio. Our results may prove helpful in the realization of experiments with an increasing number of spectral dimensions and increasingly higher order of nonlinearity.

## Supplementary material

See the supplementary material for the simulations of the sixth-order rephasing 2Q–1Q 2D spectrum with increasing electric field amplitude and experimental difference maps of the fourth-order rephasing and nonrephasing 1Q–1Q 2D spectra.

## Acknowledgements

The authors acknowledge funding by the European Research Council (ERC) within Advanced Grant No. 101141366.

## Conflict of interest

The authors declare no competing interest.

## Data availability

The data that support the findings of this study, including the Matlab scripts for the determination of the phase-cycling schemes, are openly available in WueData at https://doi.org/10.58160/stAWtPUnpVRsOspy.[90]

# Supplementary material:
# Cogwheel phase cycling in population-detected optical coherent multidimensional spectroscopy

Ajay Jayachandran[†,1], Stefan Mueller[†,1], and Tobias Brixner[1,2,3,*]

[1]*Institut für Physikalische und Theoretische Chemie, Universität Würzburg, Am Hubland, 97074 Würzburg, Germany*

[2]*Center for Nanosystems Chemistry (CNC), Universität Würzburg, Theodor-Boveri-Weg, 97074 Würzburg, Germany*

[3]*Institute for Sustainable Chemistry & Catalysis with Boron (ICB), Universität Würzburg, Am Hubland, 97074 Würzburg, Germany*

[†]*These authors contributed equally*

*Author to whom correspondence should be addressed:*

*[*tobias.brixner@uni-wuerzburg.de](mailto:tobias.brixner@uni-wuerzburg.de)*

## Contents

# SI. Simulations with increasing electric field amplitude

When increasing the amplitude of the electric field, we observe in Fig. S1(a) that the integrated sixth-order rephasing 2Q–1Q signal does not follow the same trend in 21-fold nested (NES21) and in 18-fold cogwheel phase cycling (COG18). Instead, we observe that the integrated COG18 signal increases less than the NES21 signal with increasing electric field amplitude, indicating contamination by higher nonlinear orders, as they contribute to the desired signal with opposite sign. In Fig. S1(b)–(d), we show the 2D spectra at the highest electric field amplitude of $14\times10^{-4}$. We also observe that the difference signal amplitude is higher relative to that of the NES21 and COG18 signals compared to the signal amplitudes reported in Fig. 2(c)–(e) in the main manuscript. Note that in the simulation shown in the main manuscript, we used a lower electric field amplitude of $2.25\times10^{-4}$. Moreover, the COG18 signal at the highest field amplitude appears to be slightly distorted compared to the NES21 signal [compare Figs. S1(b) and (c)], which further indicates that the COG18 signal is contaminated by higher-order signals. It is difficult to identify which higher-order signals contribute to that contamination. The presence of ripples in the difference signal [Fig. S1(d)] indicates that dynamics are involved that extend over the duration of the sampled coherence times and thus presumably correspond to population dynamics. We therefore believe that the COG18 signal could be contaminated with eighth-order signals of the type 0Q–ZQ or ZQ–0Q.

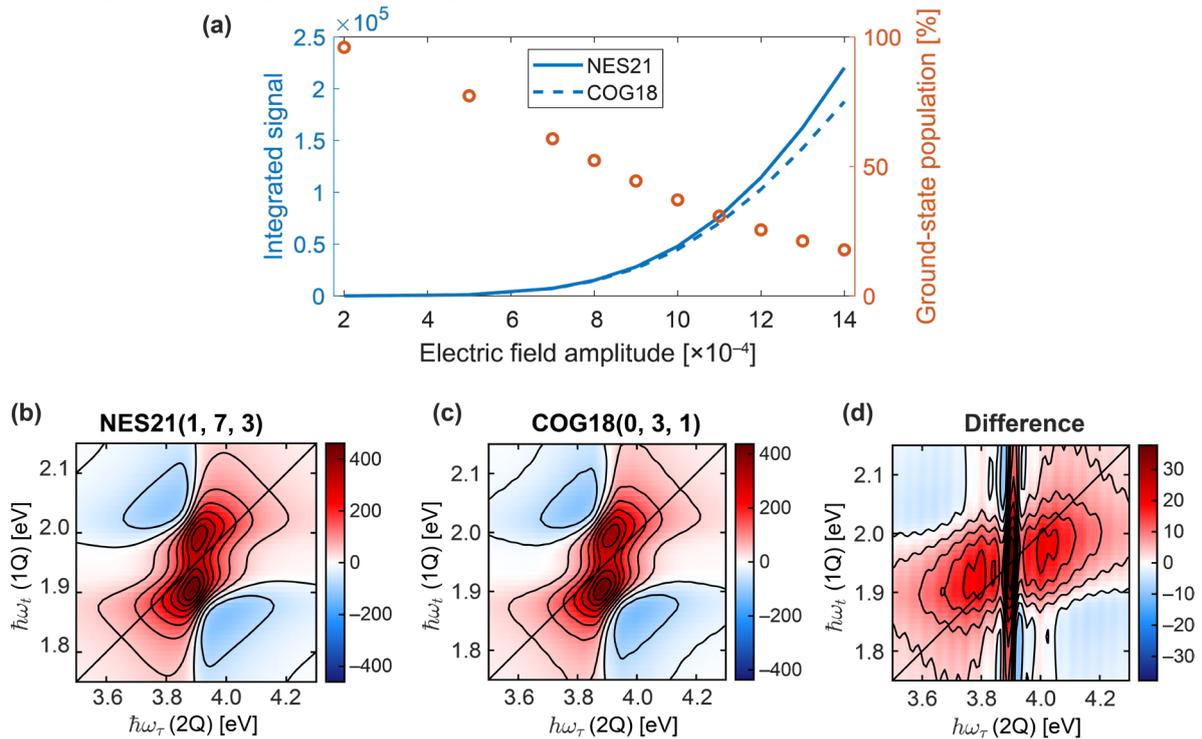

**Fig. S1.** Cogwheel phase cycling versus nested phase cycling in dependence of the electric field amplitude. (a) Integrated signal amplitudes of the sixth-order rephasing 2Q–1Q signal in case of 21-fold nested (NES21) and 18-fold cogwheel phase cycling (COG18), plotted against the amplitude of the electric field. The population of the ground state after the action of the excitation sequence with maximum pulse overlap is shown in orange. (b) Real part of the rephasing 2Q–1Q 2D spectrum in case of NES21 and (c) COG18, calculated with an electric field amplitude of $14\times10^{-4}$. (e) Difference spectrum, obtained from subtracting the spectrum of (b) from that of (c). All 2D spectra are drawn with nine linearly spaced contour lines.

It is worth noting that we have considered field amplitudes in the simulation that may go beyond usual experimental conditions [see Fig. S1(a)]. For example, at an electric field amplitude of $9\times10^{-4}$, the ground state is already more than half depopulated after the action of excitation sequence (determined at maximum temporal pulse overlap). Thus, we do not expect a manifestation of such a higher-order contamination in an experimental scenario because the ground state will not be depopulated to such an extent. The signal strength of the higher-order contaminations would be probably below the noise floor. We also confirm this hypothesis with experimental data on the example of another nonlinear signal in Sec. SII.

## SII. Experimental difference maps

In Fig. S2, we show the 2D spectra for three delays $T$ that result from subtracting the 1Q–1Q 2D spectra acquired by NES27(1, 3, 3, 3) from those acquired by COG20(0, 1, 6, 10) [Fig. S2(a)] and COG20(0, 1, 10, 6) [Fig. S2(b)]. Especially in the rephasing 1Q–1Q 2D difference spectra in Fig. S2(a), we observe only noise rather than signals that differ significantly from the noise level. Thus, no signal aliasing or higher-order contamination is introduced by cogwheel phase cycling. In case of the nonrephasing 1Q–1Q 2D spectra shown in Fig. S2(b) we observe a signal that is slightly above the noise floor. This difference signal arises due to differences in the excitation conditions between the experiments with NES27(1, 3, 3, 3) and COG20(0, 1, 10, 6). In the experiment with COG20(0, 1, 10, 6), the pulse energy was approximately 10 nJ higher than that of the experiments with NES27(1, 3, 3, 3) and COG20(0, 1, 6, 10). We thus attribute the difference signal observed in Fig. S2(b) to different fourth-order signal strengths between the two experiments rather than a higher-order contamination.

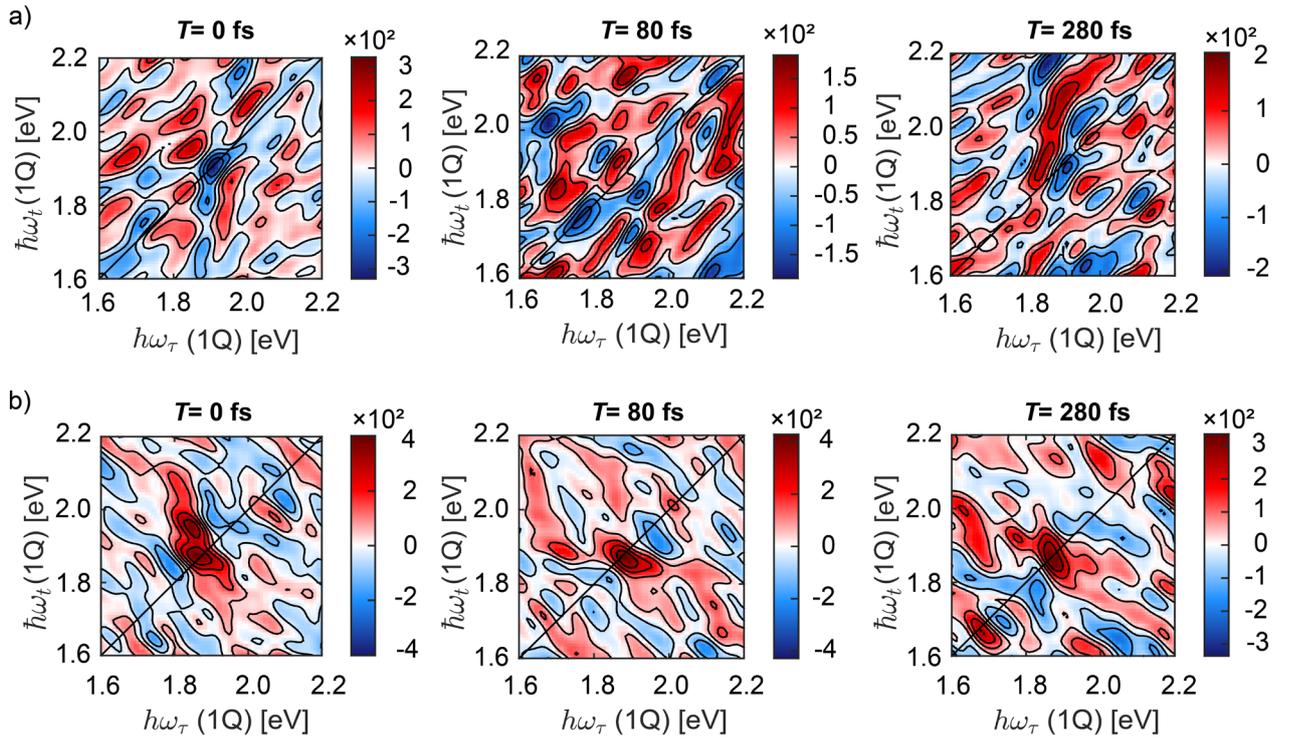

**Fig. S2.** (a) Real parts of the difference of rephasing 1Q–1Q 2D spectra of rhodamine 700 obtained for different delays $T$ by using NES27(1, 3, 3, 3) and COG20(0, 1, 6, 10) at $T$ = 0 fs, $T$ = 80 fs, and $T$ = 280 fs. (b) Real parts of the difference of nonrephasing 1Q–1Q 2D spectra of rhodamine 700 obtained for different delays $T$ by using NES27(1, 3, 3, 3) and COG20(0, 1, 10, 6) at $T$ = 0 fs, $T$ = 80 fs, and $T$ = 280 fs. All 2D spectra are drawn with six linearly spaced contour lines.